\documentclass[superscriptaddress,preprintnumbers,prl,twocolumn]{revtex4}

\usepackage{style}

\usepackage{float}

\begin{document}
  
\title{Circumventing Magnetostatic Reciprocity: a Diode for Magnetic Fields}

\author{J. Prat-Camps}
\email{j.prat.camps@gmail.com}
\affiliation{Institute for Quantum Optics and Quantum Information of the
Austrian Academy of Sciences, A-6020 Innsbruck, Austria}
\affiliation{Institute for Theoretical Physics, University of Innsbruck, A-6020 Innsbruck, Austria}
\affiliation{{INTERACT Lab, School of Engineering and Informatics, University of Sussex, Brighton BN1 9RH, UK}}

\author{P. Maurer}
\affiliation{Institute for Quantum Optics and Quantum Information of the
Austrian Academy of Sciences, A-6020 Innsbruck, Austria}
\affiliation{Institute for Theoretical Physics, University of Innsbruck, A-6020 Innsbruck, Austria}

\author{G. Kirchmair}
\affiliation{Institute for Quantum Optics and Quantum Information of the
Austrian Academy of Sciences, A-6020 Innsbruck, Austria}
\affiliation{Institute for Experimental Physics, University of Innsbruck, A-6020 Innsbruck, Austria}

\author{O. Romero-Isart}
\affiliation{Institute for Quantum Optics and Quantum Information of the
Austrian Academy of Sciences, A-6020 Innsbruck, Austria}
\affiliation{Institute for Theoretical Physics, University of Innsbruck, A-6020 Innsbruck, Austria}

\begin{abstract}
	Lorentz reciprocity establishes a stringent relation between electromagnetic fields and their sources. For static magnetic fields, a relation between magnetic sources and fields can be drawn in analogy to the Green's reciprocity principle for electrostatics. So far, the magnetostatic reciprocity principle remains unchallenged and the magnetostatic interaction is assumed to be symmetric (reciprocal). Here, we theoretically and experimentally show that a linear and isotropic electrically conductive material moving with constant velocity is able to circumvent the magnetostatic reciprocity principle and realize a diode for magnetic fields. This result is demonstrated by measuring an extremely asymmetric magnetic coupling between two coils that are located near a moving conductor. The possibility to generate controlled unidirectional magnetic couplings implies that the mutual inductances between magnetic elements or circuits can be made extremelly asymmetric. We anticipate that this result will provide novel possibilities for applications and technologies based on magnetically coupled elements and might open fundamentally new avenues in artificial magnetic spin systems.
\end{abstract}
    
\maketitle

\begin{figure}[b]
	\begin{center}
		\includegraphics[width=0.5\textwidth]{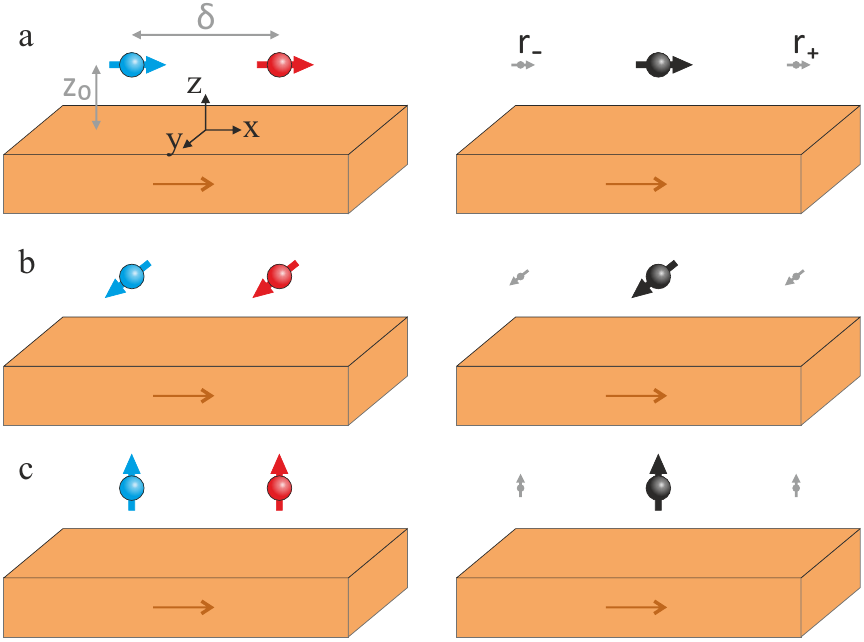}
		\caption{\footnotesize (Left) Sketch of the magnetic dipoles between which magnetic reciprocity is evaluated ($\mathbf{m}_1$ in blue and $\mathbf{m}_2$ in red). For translationally symmetric systems, this is equivalent to considering a single dipole (right) and evaluating the component of the field parallel to the dipole at $\mathbf{r}_+$ and $\mathbf{r}_-$.}
		\label{fig1}
	\end{center}
\end{figure}

Lorentz reciprocity is a general principle that relates electromagnetic (EM) fields with their sources. Arising directly from Maxwell equations, it has a fundamental importance in a huge variety of EM systems and technologies, ranging from radio-wave and microwave antennas to photonic communication systems, to name only few. Finding ways to break the Lorentz reciprocity principle has raised a lot of interest lately~{\cite{Potton2004,Caloz2018}}, since it is a necessary condition to build true EM isolators that allow the propagation of signals in one direction while preventing back-action in the opposite one~\cite{Jalas2013}.  
Recently, it has been shown that breaking Lorentz reciprocity also allows to overcome fundamental time-bandwidth limitations in resonant systems~\cite{Tsakmakidis2017}.
The concept of reciprocity extends to other physical systems, like acoustic wave propagation or mechanical systems~\cite{Fleury2014,Cummer2016,Coulais2017}. Also there, one aims at breaking reciprocity to achieve one-way signal propagation.
In the context of microwaves and photonic systems, the magneto-optical effect (Faraday rotation) has been traditionally used to break reciprocity. 
However, such effect relies on the application of an external magnetic bias, which makes it unsuitable for on-chip miniaturization and integration. This has prompted the development of a whole new generation of magnetic-free non-reciprocal devices mainly based on the application of other bias vectors which are odd under time reversal. This includes the spatio-temporal modulation of material properties to impart angular momentum bias~\cite{Sounas2013,Estep2014,Kamal2011,Sounas2017},  
linear momentum~\cite{Qin2014}, 
or commutation~\cite{Reiskarimian2016}. 
It has been realized that optomechanical coupling can also be used to induce electromagnetic nonreciprocity~\cite{Manipatruni2009}, see~\cite{Miri2017} and references therein.

In the static limit, Maxwell equations decouple and reciprocity needs to be revised. In electrostatics, Green's reciprocity~\cite{Griffiths,Jackson} relates two independent charge distributions, $\rho_1$ and $\rho_2$, with their corresponding electrostatic potentials, $V_1$ and $V_2$, via $\int\text{d}\textbf{r} \rho_1 V_2=\int\text{d}\textbf{r} \rho_2 V_1 $. For the magnetostatic case, one can do an analogous derivation. Consider two independent distributions of current densities, $\mathbf{J}_1$ and $\mathbf{J}_2$, that create the magnetic fields $\mathbf{H}_1$ and $\mathbf{H}_2$, respectively. The corresponding magnetic vector potentials, $\mathbf{A}_1$ and $\mathbf{A}_2$, are related to the fields through the magnetic permeability tensor, $\bar{\bar{\mu}}$, as $\bar{\bar{\mu}}\,\mathbf{H}_i=\nabla \times \mathbf{A}_i$ ($i=1,2$). Using the two sets of magnetostatic Maxwell equations and manipulating them, one finds {$\nabla \cdot ( \mathbf{H}_1 \times \mathbf{A}_2-\mathbf{H}_2 \times \mathbf{A}_1 )=\mathbf{H}_2 \bar{\bar{\mu}} \mathbf{H}_1-\mathbf{H}_1 \bar{\bar{\mu}} \mathbf{H}_2+\mathbf{A}_2\cdot\mathbf{J}_1-\mathbf{A}_1\cdot\mathbf{J}_2$ (see Supplemental Material~\cite{SM}).}
The first two {terms} cancel out if (i) permeability is a symmetric tensor, $\bar{\bar{\mu}}=\bar{\bar{\mu}}^{\rm T}$, and (ii) $\bar{\bar{\mu}}$ is linear (i.e. does not depend on the magnetic field). {By} integrating over all space, the left-hand-side {vanishes}. This leads to the reciprocity condition for magnetostatic fields,~\cite{Zangwill}
\begin{equation}
\int \text{d}\textbf{r}\mathbf{A}_2\cdot\mathbf{J}_1  = \int\text{d}\textbf{r} \mathbf{A}_1\cdot\mathbf{J}_2.\label{eq2}
\end{equation}
{This expression reads similar to the Lorentz reciprocity equation for electromagnetic waves and localized sources, $\int \text{d}\textbf{r}\mathbf{E}_2\cdot\mathbf{J}_1  = \int\text{d}\textbf{r} \mathbf{E}_1\cdot\mathbf{J}_2$~\cite{Potton2004,Tsakmakidis2017}, with $\mathbf{E}$ being the electric field. As shown in the Supplemental Material~\cite{SM}, though, the derivation of Lorentz reciprocity condition assumes coupled electric and magnetic fields and, thus, one cannot make the zero-frequency limit directly. 
At the same time, static conditions (no temporal variation of fields, sources or material properties) impose severe constrains when one aims at circumventing magnetostatic reciprocity; there is no magneto-optical coupling and temporal modulation of the material properties is not compatible with static conditions} \footnote{{This means that none of the electromagnetic material properties (electric permittivity, magnetic permeability, or electrical conductivity) can change over time, i.e. none of these properties can have any explicit dependence on time.}}.

{The magnetostatic reciprocity condition in} \eqnref{eq2} can be can be rewritten in different ways. When sources are point magnetic dipoles with moments $\mathbf{m}_1$ and $\mathbf{m}_2$, located at positions $\mathbf{r}_1$ and $\mathbf{r}_2$, respectively, it simplifies to $\mathbf{B}_2(\mathbf{r}_1)\cdot \mathbf{m}_1=\mathbf{B}_1(\mathbf{r}_2)\cdot \mathbf{m}_2$, where $\mathbf{B}_i$ is the magnetic induction field created by the $i$th dipole. Alternatively, when sources are closed magnetic {circuits,} \eqnref{eq2} becomes $M_{12}=M_{21}$, being $M_{nm}$ the mutual inductance between the $n$th and the $m$th circuits. This shows how the magnetic reciprocity principle is responsible for the symmetry of magnetic couplings~\cite{Tesche,Chua}.

The magnetostatic reciprocity principle formulated in \eqnref{eq2} holds for linear materials with locally symmetric permeability tensors [$\bar{\bar{\mu}}(\mathbf{r})= \bar{\bar{\mu}}(\mathbf{r})^{\rm T}$]. This includes magnetic metamaterials~\cite{Wood2007,Magnus2008,Anlage2010,antimagnet,Narayana2012,Wang2013,experimental,ML} which, despite of being complex arrangements of different magnetic materials with unusual effective magnetic properties, are locally symmetric. Hence, magnetic metamaterials cannot break the magnetic reciprocity principle even in extremely counter-intuitive cases, {see~\cite{SM}}.

Let us now show how, in spite of using linear, isotropic, and homogeneous materials one can optimally circumvent the magnetic reciprocity principle by means of a moving electrical conductor. 
When a conductor with electrical conductivity $\sigma$ moves with velocity $v\ll c$ ($c$ is the speed of light) in the presence of a magnetic field, a current density given by $\mathbf{J}_{\rm mc}=\sigma \mathbf{v}\times\mathbf{B}$ is induced~\cite{Landau}. If one includes this term in the previous {reciprocity} derivation, an extra factor {appears} reading $\sigma \mathbf{v}\cdot[(\nabla \times \mathbf{A}_1)\times \mathbf{A}_2-(\nabla \times \mathbf{A}_2)\times \mathbf{A}_1]$. This factor, {generally} different from zero, shows how a moving conductor can break reciprocity.

As a particular case, we consider a semi-infinite conductor that extends to $z<0$. We assume it has a constant electrical conductivity $\sigma$ and a velocity $\mathbf{v}=v {\rm \hat{e}}_x$. We evaluate the magnetic reciprocity between two identical dipoles, $\mathbf{m}_1=\mathbf{m}_2=m\,{\rm \hat{e}}_j$, (being $\hat{e}_j$ a unit vector, $j=x,y,z$) situated at $\mathbf{r}_1=(-\delta/2,0,z_0)$ and $\mathbf{r}_2=(\delta/2,0,z_0)$, respectively, see Fig.~\ref{fig1} left ($\delta,z_0>0$). Since the conductor is translationaly invariant along x, this problem is equivalent to considering a single magnetic point dipole with moment $\mathbf{m}=m\,{\rm \hat{e}}_j$ located at $\mathbf{r}=(0,0,z_0)$ and evaluating the magnetic field at the positions $\mathbf{r}_+=(\delta,0,z_0)$ and $\mathbf{r}_-=(-\delta,0,z_0)$, see Fig.~\ref{fig1} right. Reciprocity dictates that the \textit{isolation}, defined as ${\cal I}_j \equiv B_j(\mathbf{r}_-)/B_j(\mathbf{r}_+)$, is ${\cal I}_j=1$. 

\begin{figure}[b]
	\begin{center}
		\includegraphics[width=0.5\textwidth]{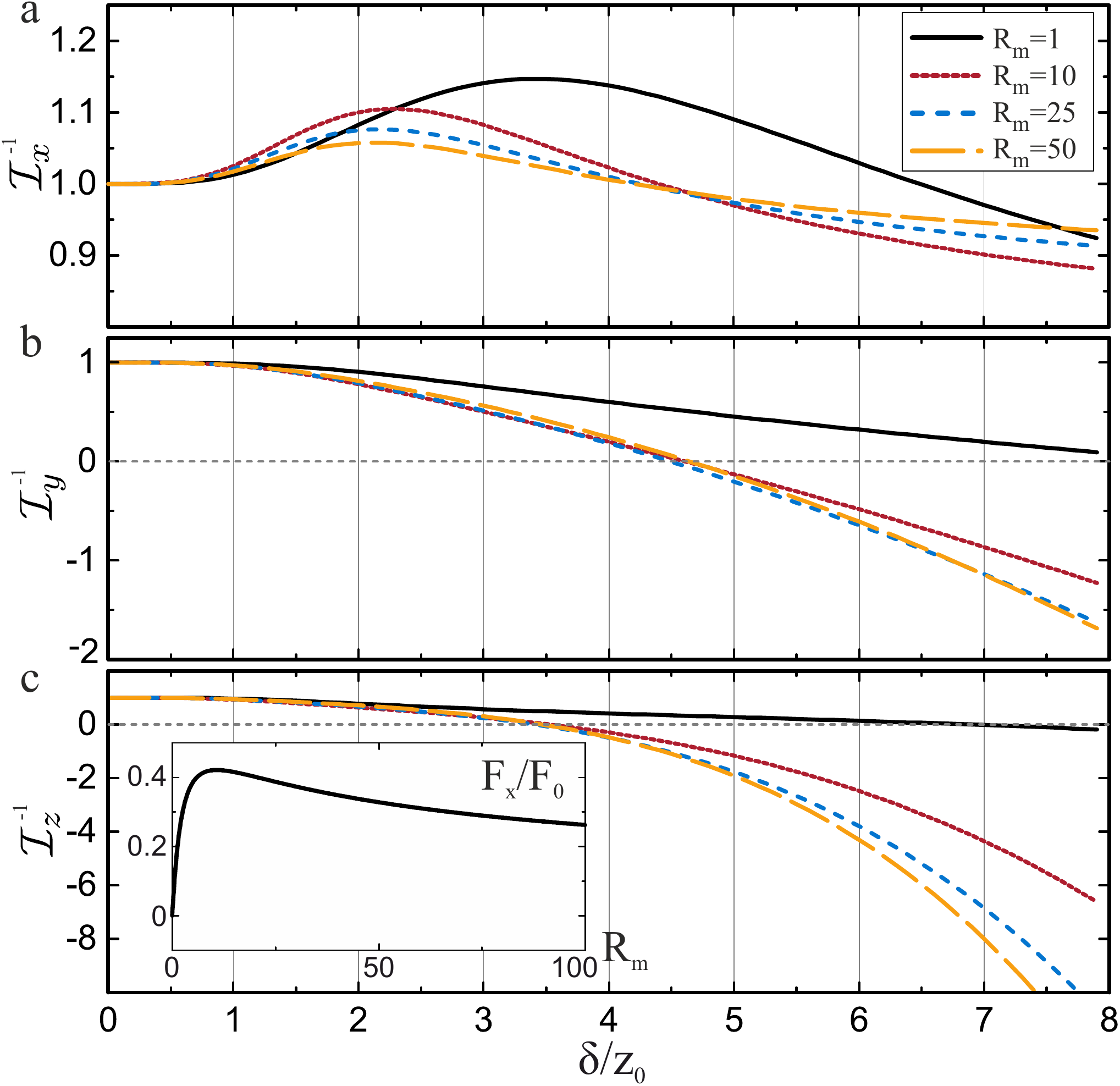}
		\caption{\footnotesize Plots of the inverse of the isolation as a function of $\delta/z_0$ for dipoles oriented along (a) x, (b) y, and (c) z-directions for different values of $R_{\rm m}$. Inset in (c) shows the normalized force $F_x/F_0$ [$F_0\equiv \mu_0 m^2/(8\pi^2 z_0^4)$] on a z-oriented dipole as a function of $R_{\rm m}$. Notice that, since the materials involved are linear, these plots (and therefore, the points of infinite isolation) do not depend on the modulus of the magnetic dipole.}
		\label{fig2}
	\end{center}
\end{figure}

We analytically solve the Lorentz-transformed problem of a dipole moving with constant velocity at a fixed height $z_0$ above a semi-infinite surface characterized by a complex permitivity {$\varepsilon(\omega)=1+\im \sigma/(\varepsilon_0\omega)$~\cite{SM}}. The scattered field is obtained everywhere in the upper half-space. 
In general, this field does not show any clear symmetry and strongly depends on the magnetic Reynolds number, $R_{\rm m} \equiv \mu_0 \sigma v z_0$ (being $\mu_0$ the vacuum permeability). For small Reynolds numbers ($R_{\rm m}\ll 1$), the scattered field can be approximated to an anti-symmetric function of $\delta$. In this case the effect of the conductor is clear; since the field of the bare dipole is symmetric, the moving conductor increases the field on one side but decreases it on the other.
The inverse of the isolation between dipoles, arranged in the three different configurations, is plotted in Fig.~\ref{fig2}; the curves only depend on $\delta/z_0$ and $R_{\rm m}$.
These plots show how the moving conductor generates isolations different from 1 and, thus, breaks magnetic reciprocity for the three different dipole orientations. However, while isolations between x-oriented dipoles are small (values near one), y and z-orientations result in isolations that go from positive to negative values through a divergence. 
The existence of $\delta$'s for which the isolation is infinite (we refer to the points where ${\cal I}_j^{-1}=0$ as $\delta_0^j$ for $j=y,z$) demonstrates that one can achieve a maximally asymmetric (unidirectional) magnetic coupling between the dipoles. 
For example, for two dipoles oriented along z and located at $\mathbf{r}_1=(-\delta_0^z/2,0,z_0)$ and $\mathbf{r}_2=(\delta_0^z/2,0,z_0)$, one finds that $B_{z,1}(\mathbf{r}_2)=0$ whilst $B_{z,2}(\mathbf{r}_1)\neq0$.
If dipoles are interpreted as small circular coils with axis along the z-direction, then this means that the magnetic flux threading coil 1 is different from zero whilst the flux through coil 2 is zero. Therefore, the mutual inductance between the two magnetic elements becomes maximally asymmetric, with $M_{12}=0$ and $M_{21}\neq0$. 
In this sense, an unidirectional magnetic coupling is achieved, realizing a perfect \textit{diode for magnetic fields}.

\begin{figure}[b!]
	\begin{center}
		\includegraphics[width=0.50\textwidth]{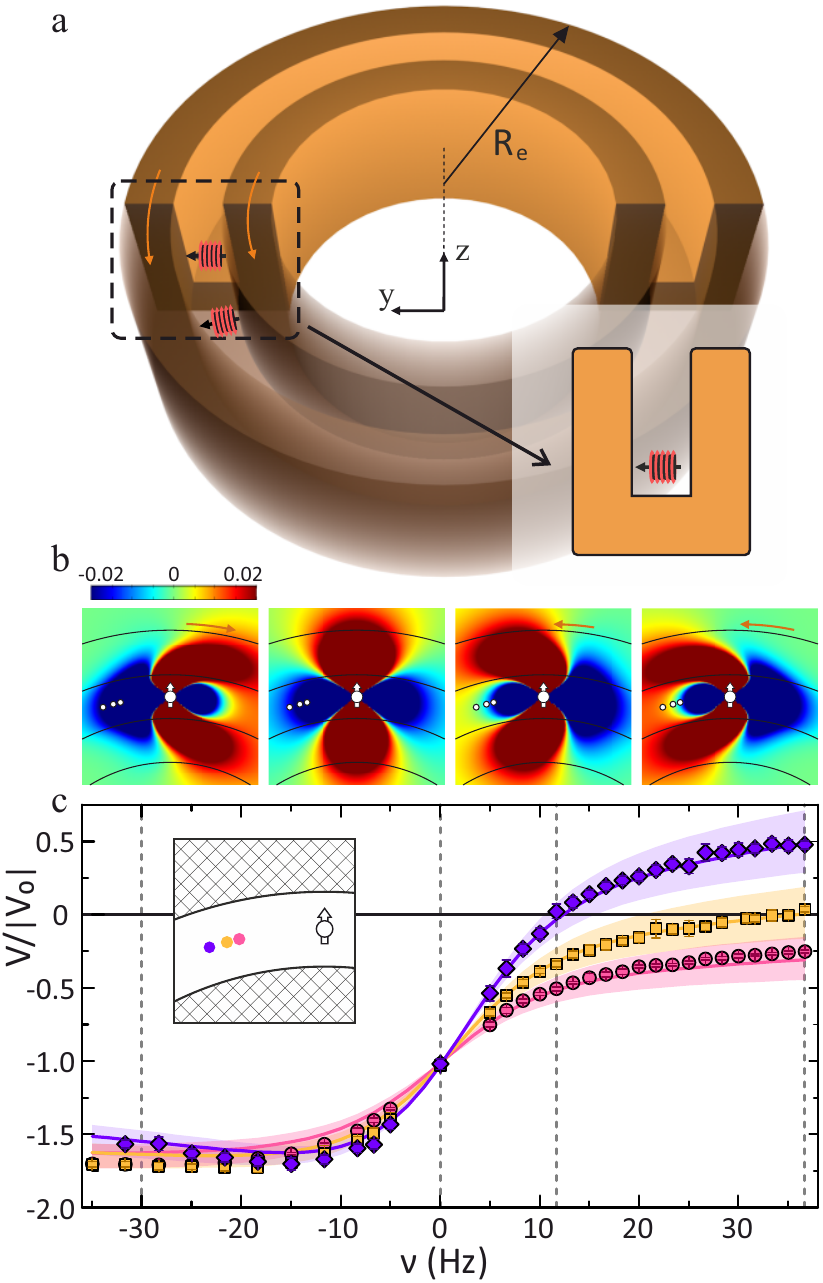}
		\caption{\linespread{1.0} \scriptsize (a) Sketch of the experimental setup; a circular U-shaped conductor (orange, with $R_e=65$mm) moves with rotation frequency $\nu$ around the z-axis (arrows indicate the positive rotation direction). Two coils (in red), whose axis are radially aligned, are used to measure the magnetic coupling between them. 
			(b) Numerical calculations for an oscillating magnetic dipole (in white, for $\omega_0/(2\pi)=9$Hz). Colors correspond to the real part of the normalized radial field, $B_{\rho}/B_0$, (where $B_0\equiv \mu_0 m/(2\pi z_0^3)$ with $z_0=5$mm) for different rotation frequencies of the conductor, $\nu=-30, \, 0, \, 11.7,$ and $36.7$Hz (from left to right). Plots show the magnetic field distribution evaluated at the plane of the dipole. White dots indicate positions where measurements were taken. 
			(c) Out-of-phase component of the voltage measured in the receiving coil (symbols) as a function of the velocity of the conductor (for a signal frequency of $\omega_0/(2\pi)=9$Hz). Measurements were taken at three different distances from the source coil, $r_1=11.4$mm (pink), $r_2=13.1$mm (yellow), and $r_3=15.5$mm (purple), see inset. For each distance, measurements are normalized to the voltage induced at the receiving coil in free space, $|V_0|$. 
			Solid lines are the corresponding numerical calculations considering point dipoles. Shadow areas are defined by considering uncertainties in the experimental parameters used for the numerical {calculations}~\cite{SM}. Dashed vertical lines indicate the frequencies of the numerical calculations in b. {Error bars (1 sigma) are shown for the three cases; most of them are symbol-size or smaller}.
		}
		\label{fig3}
	\end{center}
\end{figure}

Note that this mechanism is intrinsically lossy; one needs to add energy to the system in order to keep the conductor moving at constant velocity and overcome the magnetic friction originating from the induced eddy currents. The power dissipated by the system of Fig.~\ref{fig1} right is given by $P=-v F_x$, where $F_x$ is the x-component of the force acting on the dipole as a result of these currents. The force can be analytically calculated from the field scattered by the {conductor~\cite{SM}}. 
As shown in the inset of Fig.~\ref{fig2}c, the normalized force only depends on $R_{\rm m}$ and has a non-monotonic behaviour; it is 0 for $R_{\rm m}=0$ (when the conductor is at rest), grows linearly for small $R_{\rm m}$, reaches a maximum value for $R_{\rm m}\approx 10$, and decreases as $R_{\rm}^{-1/2}$ for $R_{\rm m}\gg 10$. 
This force has a similar velocity dependence as the vacuum frictional force between two conducting surfaces~\cite{Pendry1997}, which is maximal for a certain velocity and monotonically decreases for bigger values.
Interestingly, it can be demonstrated that for a perfect electric conductor ($\varepsilon \rightarrow \infty$), reciprocity is preserved and ${\cal I}=1$ for all $\delta$. In this ideal case the system is lossless and the dipole experiences no force. 
For consistency, we also checked that the Lorentz-transformed problem with the dipole at rest and the conductor moving with constant velocity leads to the same results. We solved this problem numerically with COMSOL Multiphysics by introducing a free current density $\mathbf{J}_{\rm mc}$ in the conductor, finding good agreement with our analytical results.

Finally, we remark that these results are also valid for low-frequency oscillating magnetic fields. {We analytically} solve the problem of a z-oriented magnetic dipole, whose moment oscillates as $\mathbf{m}(t)=m \cos(\omega_0 t)\hat{\rm e}_z$. For $\omega_0 \ll |v/z_0|$, one finds that the magnetic field distribution is the same as for the static case, simply modulated by a $\cos(\omega_0 t)$ function~\cite{SM}. Therefore, even for low-frequency oscillating magnetic sources and circuits, the moving conductor is able to generate a maximally asymmetric magnetic coupling between them.

We shall now present the experimental demonstration of these results. Our setup consists of a circularly symmetric conductor with a U-shaped cross-section, as sketched in Fig.~\ref{fig3}a, that moves with constant angular velocity around its axial symmetry axis. The previous analysis indicates that the magnetic moment of the dipole has to be perpendicular to the velocity in order to generate points of infinite isolation. 
For this reason, we put a small coil inside the moving conductor space, with its axis pointing along the radial direction. A second coil is placed at a given distance with analogous radial orientation, and the magnetic coupling between them {is measured~\cite{SM}}. 
{For experimental convenience, the experiment is performed with low-frequency oscillating magnetic fields.}
We use a signal generator to feed the first coil, while the voltage induced in the second (pick-up) coil is measured through a lock-in amplifier. 
{Lock-in measurements provide a good signal-to-noise ratio even for small magnetic fields and allow to get rid of slowly fluctuating magnetic fields in the environment. At the low frequencies we consider, the coupling between magnetic and induced electric field is negligible and, thus, these measurements effectively describe the static case.
}
Measurements of the out-of-phase voltage for a signal frequency of $\omega_0/(2\pi)=9$Hz are shown in Fig.~\ref{fig3}c, as a function of the rotation frequency of the conductor,~$\nu$. Measurements are repeated for three different positions of the pick-up coil (see inset). 
For positive rotation frequencies, the measured voltage decreases and crosses 0 for positions $r_2$ and $r_3$ of the pick-up coil. At position $r_1$, the field scattered by the conductor is not able to fully cancel the field of the source for the velocities we considered. When moving in the opposite direction, the conductor increases the measured voltage. These measurements convincingly demonstrate that magnetic reciprocity is broken and that points of infinite isolation (for which the measured voltage is zero for positive rotation frequency but different from zero for negative) are generated by means of a moving conductor. These zero-voltage points are found in spite of the extended size of the pick-up coil; the field goes from positive to negative values around the zero and thus, the total magnetic flux threading the coil cancels out at some point.
{As can be seen, the error bars associated to our measurements are very small compared to the measured voltages. These errors come from the measured voltage fluctuations over time (plotted error bars correspond to 1 sigma).}

All these measurements agree very well with the corresponding 3D numerical calculations (solid lines in Fig.~\ref{fig3}c) considering the coils as point dipoles. 
{The main source of uncertainty between our measurements and the numerical calculations comes from the positioning of the coils relative to each other and to the moving conductor. We tried to estimate the effects of imprecise positioning by running different numerical calculations in which we changed the distance between the coils ($\pm0.5$mm) and their relative position with respect to the conductor ($\pm0.5$mm in $z$-direction). The results were used to create the shadow bands in Fig.~\ref{fig3}c, defined as the result with the largest deviation for each $\nu$ from the nominal calculation.}
{Numerical calculations also provide} a deeper understanding on how the conductor shapes the distribution of magnetic field. In Fig.~\ref{fig3}b, we show numerical calculations of the real part of the $B_{\rho}$ field (being $B_{\rho}$ the radial component of the field in cylindrical coordinates, $\rho=\sqrt{x^2+y^2}$) created by a magnetic dipole (in white) oscillating at a frequency $\omega_0/(2\pi)=9$Hz. The symmetric field distribution when the conductor is at rest (second panel) becomes clearly asymmetric as it moves in one direction (1st panel). When moving in the opposite direction (3rd and 4th panels), the field distribution flips direction. This evidences the existence of points of infinite isolation (points of zero field, in green color). 

Measurements were repeated for higher signal {frequencies~\cite{SM}}. In all cases the agreement with the corresponding numerical calculations is excellent.  
Finally, we measure the actual mutual inductance between the two coils to demonstrate how extremely asymmetric values are achieved. The second coil is placed at $r_2$ and is connected to the lock-in amplifier, while the first coil is connected to the signal generator. With the conductor at rest ($\nu=0$), we measure $M_{12}=(22+3\im)$nH. We then exchange the connections to the coils {and measure} the opposite coupling, finding a symmetric mutual inductance, $M_{21}=M_{12}$, in agreement with the magnetic reciprocity principle. This same procedure is repeated with the conductor moving at $\nu=33.3$Hz. In this case, we first measure $M_{12}'=(0+2\im)$nH and, after exchanging the connections, we find $M_{21}'=(36+0\im)$nH~\footnote{In all these measurements the imaginary part of the mutual inductance results from the eddy-current losses in the conductor due to the fact that measurements are done at a finite frequency (9Hz). In the strict static case, inductances would be purely real.} 
(both measurements have an {error of $\pm0.6$nH}~\cite{SM}). 
These measurements demonstrate how, by tuning the velocity of the conductor, the magnetic coupling between the coils becomes unidirectional.

The use of a moving conductive material to break magnetic reciprocity boils down to the Lorentz force that the free electrons of the conductor experience as they move through the magnetic field. In principle, one could replace the mechanical movement of the whole material by an externally applied electric field, which would force the electrons to move with a constant mean velocity in the conductor according to Ohm's law. While theoretically correct, this approach is limited by the small mean velocity at which electrons move {in metals} for reasonable current densities. For copper, for example, the standard maximum current density of 500A/cm$^2$ corresponds to mean velocities $\sim 4\times10^{-4}$m/s, in contrast to the linear velocities achieved in our setup of $\sim 3.1$m/s for $\nu=10$Hz {({see~\cite{SM}} for a detailed discussion using the Drude model)}. 
{Interestingly, other materials like graphene exhibit carrier mobilities that can be more than three orders of magnitude larger than in copper~\cite{Chen2008} while being able to sustain  current densities on the order of $\sim 10^8$A/cm$^2$~\cite{Muralia2009}. Hence, graphene is an interesting candidate to explore implementations that do not rely on mechanical movement of {macroscopic objects}.
}

In conclusion, we have demonstrated that the magnetostatic reciprocity principle can be circumvented by means of a linear and isotropic electrical conductor moving with constant velocity. The non-reciprocal response of the system is controlled trough the velocity of the conductor, making it possible to achieve an infinite magnetic isolation (i.e. a perfectly unidirectional magnetic coupling) and to realize a diode for magnetic fields. The concept, which relies only on linear materials and low (non-relativistic) velocities, may open the door to novel possibilities for a large number of systems and technologies that employ magnetically coupled elements. 
{In particular, the breaking of magnetostatic reciprocity could be useful to increase the efficiency of magnetically-based wireless power transfer technologies. This would allow the energy to flow from the emitting to the receiving circuit but would prevent the flow in the opposite direction. Other key technologies based on magnetically coupled circuits, like transformers, could also benefit from this same principle.}
Results presented here could also open new horizons in fundamental research areas, like artificial magnetic spin systems. A conductor moving near a system of artificial spins would alter the reciprocal dipole-dipole interaction between them, potentially forcing the system to crystallize in non-conventional structures. 

\vspace{1mm}
This work is supported by the European Research Council (ERC-2013-StG 335489 QSuperMag) and the Austrian Federal Ministry of Science, Research, and Economy (BMWFW). JPC acknowledges discussions with the Superconductivity Group of the Universitat Autonoma de Barcelona. We acknowledge the help of S. Oleschko and our in-house workshop for the fabrication of the experimental setup.

\end{document}


\title{SUPPLEMENTAL MATERIAL \\Circumventing Magnetostatic Reciprocity: a Diode for Magnetic Fields}

\author{J. Prat-Camps, P. Maurer, G. Kirchmair, O. Romero-Isart}

    
    
\maketitle

\tableofcontents

\section{Reciprocity for static magnetic fields and sources}

{
Let us derive a reciprocity condition for static magnetic fields. We consider two independent current densities, $\mathbf{J}^{\rm f}_j$, that give rise to two independent sets of magnetic fields. These fields fulfill the two magnetostatic Maxwell equations,
\begin{align}
\nabla \times \mathbf{H}_j&=\mathbf{J}^{\rm f}_j,\label{eq1}\\
\nabla \cdot \mathbf{B}_j&=0,\label{eq2}
\end{align}
where $j=1,2$ stands for the two independent distributions. \eqnref{eq2} is fulfilled if we define a magnetic vector potential, $\mathbf{A}$, such that
\begin{align}
\nabla \times \mathbf{A}_j&= \bar{\bar{\mu}}\mathbf{H}_j,\label{eq4}
\end{align}
where we have used the constitutive relation $\mathbf{B}_j=\bar{\bar{\mu}}\mathbf{H}_j$, being $\bar{\bar{\mu}}$ the total magnetic permeability tensor. Now considering Eq.~(\ref{eq1},\ref{eq4}) for $j=1$, multiplying them from the left by $\mathbf{A}_2$ and $\mathbf{H}_2$, respectively, and adding them one obtains
\begin{align}
\mathbf{A}_2\cdot \left(\nabla \times \mathbf{H}_1\right)+\mathbf{H}_2\cdot \left(\nabla \times \mathbf{A}_1\right)=&\mathbf{H}_2 \bar{\bar{\mu}} \mathbf{H}_1 + 
\mathbf{A}_2\cdot\mathbf{J}^{\rm f}_1.\label{eq5}
\end{align}
An analogous expression is obtained starting with Eqs.~(\ref{eq1},\ref{eq4}) for $j=2$ and multiplying by the terms with $j=1$. Substracting the two equations and simplifying, one finally obtains
\begin{equation}
\begin{split}
\nabla \cdot ( \mathbf{H}_1 \times \mathbf{A}_2-\mathbf{H}_2 \times \mathbf{A}_1 )=\, 
&\mathbf{H}_2 \bar{\bar{\mu}} \mathbf{H}_1-\mathbf{H}_1 \bar{\bar{\mu}} \mathbf{H}_2 \\
&+\mathbf{A}_2\cdot\mathbf{J}^{\rm f}_1-\mathbf{A}_1\cdot\mathbf{J}^{\rm f}_2.\label{eq6}
\end{split}
\end{equation}
By integrating \eqnref{eq6} over all space, the left hand side of the equation becomes a surface integral. This integral is also equal to zero because there is no electromagnetic induction and fields decay as $\mathbf{H}\times \mathbf{A}\propto r^{-5}$.

The right hand side of the equation needs to be discussed more carefully. The first two terms cancel out if the following conditions are fulfilled: (i) permeability is a symmetric tensor $\bar{\bar{\mu}} = \bar{\bar{\mu}}^{\rm T}$, (ii) permeability is linear (i.e. does not depend on the field), and (iii) permeability does not depend on time (otherwise the previous development does not hold). Then the magnetostatic reciprocity equation reduces to
\begin{equation}
\boxed{\int_{\mathds{R}^3}\text{d}\textbf{r} \mathbf{A}_2\cdot\mathbf{J}^{\rm f}_1  = \int_{\mathds{R}^3}\text{d}\textbf{r} \mathbf{A}_1\cdot\mathbf{J}^{\rm f}_2.}\label{eqR1}
\end{equation}
}
When the sources of magnetic field are magnetized bodies, one can use $\mathbf{J}^{\rm f}_j=\nabla \times \mathbf{M}_j$ to express \eqnref{eqR1} as
\begin{equation}
\int_{\mathds{R}^3} \text{d}\textbf{r}\mathbf{B}_2\cdot\mathbf{M}_1 = \int_{\mathds{R}^3} \text{d}\textbf{r}\mathbf{B}_1\cdot\mathbf{M}_2, \label{eqR2}
\end{equation}
which reduces to 
\begin{equation}
\mathbf{B}_2(\mathbf{r}_1)\cdot \mathbf{m}_1=\mathbf{B}_1(\mathbf{r}_2)\cdot \mathbf{m}_2, \label{eqR3}
\end{equation}
for point magnetic dipoles with moments $\mathbf{m}_1$ and $\mathbf{m}_2$ located at positions $\mathbf{r}_1$ and $\mathbf{r}_2$, respectively.  
It is worth to remark that  magnetostatic reciprocity is completely analogous to the Green's reciprocity typically formulated in electrostatics as $\int \text{d}\textbf{r}\rho_1 V_2 =\int \text{d}\textbf{r}\rho_2 V_1 $ (where $\rho_j$ are two independent charge distributions and $V_j$ the corresponding electrostatic potentials)~\cite{Griffiths}.

When the sources of field are current carrying circuits, one can rewrite \eqnref{eqR1} in terms of more familiar magnitudes. Using the expression of the magnetic vector potential created by a distribution of currents $\mathbf{A}=\mu_0/(4\pi)\int\text{d}\textbf{r}' \mathbf{J}(\mathbf{r}')/|\mathbf{r}-\mathbf{r'}|$ \eqnref{eqR1} reads
\begin{align}
\frac{\mu_0}{4\pi}\int \int \text{d}\textbf{r}\,\text{d}\textbf{r}'\frac{\mathbf{J}^{\rm f}_2(\mathbf{r}')\cdot\mathbf{J}^{\rm f}_1(\mathbf{r})}{|\mathbf{r}-\mathbf{r'}|}  = 
\frac{\mu_0}{4\pi} \int \int \text{d}\textbf{r}\,\text{d}\textbf{r}'\frac{\mathbf{J}^{\rm f}_1(\mathbf{r}')\cdot\mathbf{J}^{\rm f}_2 (\mathbf{r})}{|\mathbf{r}-\mathbf{r'}|}.
\end{align}
These integrals can be identified as the mutual inductance coefficients $M_{ab}$ between the two circuits~\cite{Jackson}, and this equation can be written as
\begin{equation}
I_1 I_2 M_{21}=I_1 I_2 M_{12},
\end{equation}
where $I_j$ is the total intensity carried by circuit $j$. This final statement shows the practical relevance of magnetic reciprocity; it is implicitly present in most electro-magnetic devices that work with coupled circuits and where symmetric mutual inductances ($M_{ab}=M_{ba}$) and couplings are typically assumed. 

{
\subsection{Lorentz reciprocity for localized sources}
In the spectral domain Maxwell's equations read
\begin{align}
\nabla\cdot \mathbf{D}_j&=\rho_j^\text{f},\\
\nabla\cdot \mathbf{B}_j&=0,\\
\nabla\times\mathbf{E}_j&=\im\omega \mathbf{B}_j,\\
\nabla\times\mathbf{H}_j&=\mathbf{J}_j^\text{f}-\im\omega \mathbf{D}_j.
\end{align}
Using well known vector calculus identities and Maxwell's equations we immediately get
\begin{align}
\nabla\cdot (\mathbf{E}_k\times\mathbf{H}_l)&=\mathbf{H}_l\cdot \nabla\times\mathbf{E}_k-\mathbf{E}_k\cdot \nabla\times\mathbf{H}_l\\
&=\im \omega \mathbf{H}_l\cdot \mathbf{B}_k+\im \omega\mathbf{E}_k\cdot\mathbf{D}_l-\mathbf{E}_k\cdot\mathbf{J}_l^\text{f}.
\end{align}
After integrating this last equation over all space, using $\mathbf{D}=\bar{\bar{\epsilon}}\mathbf{E}$, $\mathbf{B}=\bar{\bar{\mu}}\mathbf{H}$ and the divergence theorem we have
\begin{equation}
\label{im}
\int_{\mathds{S}^2}\text{dS}\, \mathbf{n}\cdot (\mathbf{E}_k\times\mathbf{H}_l)=\int_{\mathds{R}^3}\text{d}\rr\,\im \omega \mathbf{H}_l\cdot \bar{\bar{\mu}}\mathbf{H}_k+\im \omega\mathbf{E}_k\cdot\bar{\bar{\epsilon}}\mathbf{E}_l-\mathbf{E}_k\cdot\mathbf{J}_l^\text{f}.
\end{equation}
Here $\bar{\bar\mu}$ and $\bar{\bar{\epsilon}}$ denote the total permeability and permittivity tensors, respectively.
Subtracting \eqnref{im} for $k=2, l=1$ from the same equation for $k=1, l=2$ leads to
\begin{equation}
\int_{\mathds{S}^2}\text{dS}\, \mathbf{n}\cdot (\mathbf{E}_1\times\mathbf{H}_2-\mathbf{E}_2\times\mathbf{H}_1)=\int_{\mathds{R}^3}\text{d}\rr\, \mathbf{E}_2\cdot\mathbf{J}_1^\text{f}-\mathbf{E}_1\cdot\mathbf{J}_2^\text{f},
\end{equation}
if the permeability and permittivity tensors fulfill
\begin{align}
\label{1}
\im\omega \mathbf{E}_1\cdot\bar{\bar{\epsilon}}\mathbf{E}_2&=\im\omega \mathbf{E}_2\cdot\bar{\bar{\epsilon}}\mathbf{E}_1,\\
\label{2}
\im \omega \mathbf{H}_2\cdot \bar{\bar{\mu}}\mathbf{H}_1&=\im \omega \mathbf{H}_1\cdot \bar{\bar{\mu}}\mathbf{H}_2.
\end{align}
For localized sources we therefore get
\begin{equation}
\int_{\mathds{R}^3}\text{d}\rr\, \left(\mathbf{E}_2\cdot\mathbf{J}_1^\text{f}-\mathbf{E}_1\cdot\mathbf{J}_2^\text{f}\right)=0. \label{eqfinal1}
\end{equation}
Let us now rewrite this theorem in terms of the magnetic vector potential and the scalar electric potential. The second Maxwell equation is fulfilled if we define a vector potential such that $\mathbf{B}_j=\nabla \times \mathbf{A}_j$. Consequently the third Maxwell equation leads to $\nabla\times \mathbf{E}_j= \im \omega \nabla\times \mathbf{A}_j$. Therefore, we have $\mathbf{E}_j=\im \omega \mathbf{A}_j-\nabla \phi_j$, where $\phi_j$ is the electric potential. Substituting this equation into the first Maxwell equations leads to $\nabla\cdot(\bar{\bar{\epsilon}}_j\cdot\nabla\phi_j) = -\rho_j^\text{f}$ in the Coulomb gauge $\nabla\cdot (\bar{\bar{\epsilon}}\mathbf{A}_j)=0$. Therefore, after using the continuity equation $\nabla\cdot\mathbf{J}_j^\text{f}=\im\omega \rho_j^\text{f}$, we immediately get
\begin{align}
 \int_{\mathds{R}^3}\text{d}\rr\, \mathbf{A}_2\cdot\mathbf{J}_1^\text{f}-\mathbf{A}_1\cdot\mathbf{J}_2^\text{f}&=\int_{\mathds{R}^3}\text{d}\rr\,  \phi_2 \rho_1^\text{f}- \phi_1 \rho_2^\text{f},\label{eqfinal2}
\end{align}
for $\omega\neq 0$. \eqnref{eqfinal2} explicitly shows how the Lorentz reciprocity condition for localized sources, typically written as in \eqnref{eqfinal1}, relates electric and magnetic quantities. Notably, the left-hand-side of this equation can be identified as part of the magnetostatic reciprocity condition, \eqnref{eqR1}, whilst the right-hand-side of the equation is part of the Green's-electrostatic reciprocity condition.


\vspace{3mm}
In the static limit, \eqnref{eqfinal2} is not valid since we explicitly used coupled electric and magnetic fields to derive it. Differently, \eqnref{eqfinal2}, is still valid for $\omega=0$. In that case, though, this equation results in a trivial identity. Rewriting Maxwell's equations for $\omega=0$ one finds 
\begin{align}
\nabla\cdot \mathbf{D}_j&=\rho_j^\text{f},\\
\nabla\cdot \mathbf{B}_j&=0,\\
\nabla\times\mathbf{E}_j&=0,\\
\nabla\times\mathbf{H}_j&=\mathbf{J}_j^\text{f},
\end{align}
where electric and magnetic fields are now explicitly decoupled. Using vector calculus identities we now find
\begin{align}
\nabla\cdot (\mathbf{E}_k\times\mathbf{H}_l)&=\mathbf{H}_l\cdot \nabla\times\mathbf{E}_k-\mathbf{E}_k\cdot \nabla\times\mathbf{H}_l=-\mathbf{E}_k\cdot\mathbf{J}_l^\text{f},
\end{align}
and integrating this equation over all space as before we find
\begin{equation}
\int_{\mathds{S}^2}\text{dS}\, \mathbf{n}\cdot (\mathbf{E}_k\times\mathbf{H}_l)=-\int_{\mathds{R}^3}\text{d}\rr\,\mathbf{E}_k\cdot\mathbf{J}_l^\text{f}.
\end{equation}
In the static limit, the left-hand-side of this equation is zero for localized electrostatic and magnetostatic fields. Since there is no electromagnetic induction, electrostatic and magnetostatic fields decay with the distance as $\mathbf{E}\times \mathbf{H}\propto r^{-5}$ and, therefore, the surface integral at infinity vanishes. As a result, each of the two terms on the left-hand-side of \eqnref{eqfinal1} are zero and the equation reduces to a trivial identity.

}

\newpage
\section{Breaking magnetostatic reciprocity}

\subsection{Non-symmetric or non-linear magnetic materials}

\begin{figure}[htb]
\begin{center}
\includegraphics[width=0.95\textwidth]{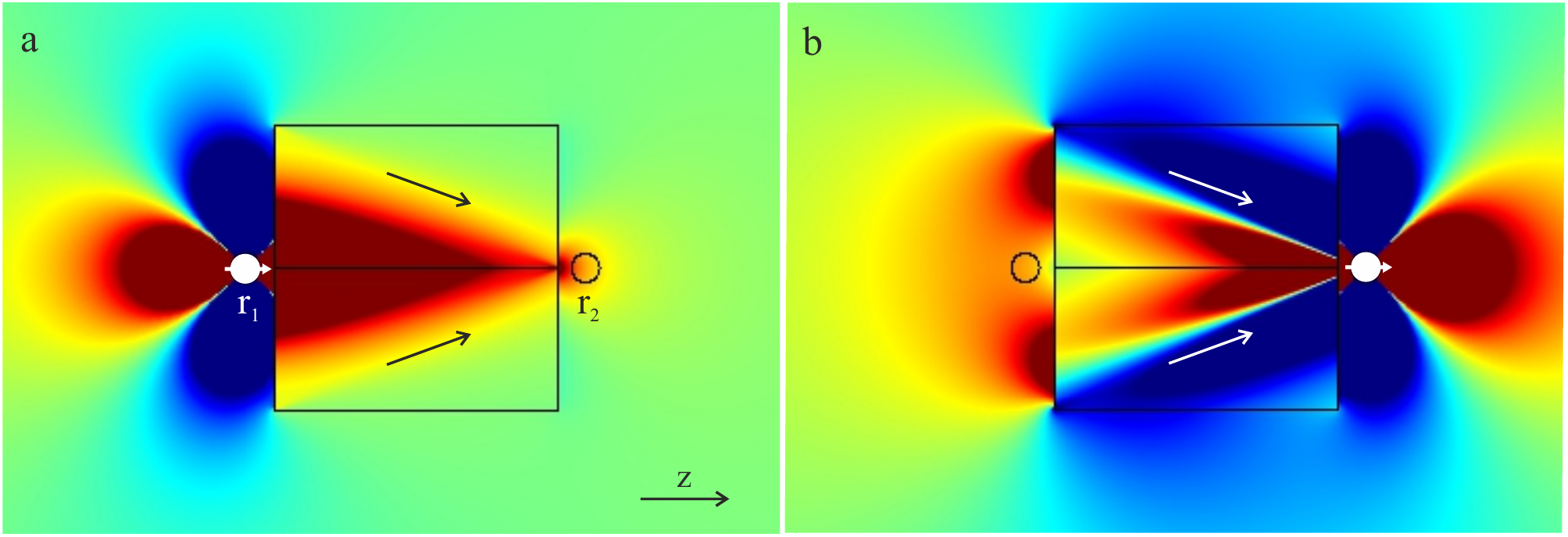}
\caption{Numerical calculations of magnetic field distributions ($B_z$ in colors, arbitrary units). The material consists of two pieces of homogeneous anisotropic magnetic material with relative permeabilities $\mu=6$ along the direction indicated by the arrows and $\mu=1/6$ in the perpendicular one. (a) Source of field is placed at $r_1$. (b) Source is placed at $r_2$.}
\label{fig1}
\end{center}
\end{figure}

Based on the derivation in the previous section, there are different conditions that would allow to break the magnetic reciprocity principle. 
The first one involves the use of a material with an asymmetric permeability tensor. Although natural magnetic materials always exhibit symmetric permeabilities, the combination of materials with different properties has made possible to create magnetic materials with exotic effective permeabilities (often called magnetic metamaterials). One could wonder whether an appropriate combination of different materials (with structures that do not have spatial symmetry, for example) could give rise to an effective asymmetric $\bar{\bar{\mu}}$ and, thus, break magnetic reciprocity. 
Unfortunately, this strategy can be disregarded. If the constituent magnetic materials are locally symmetric [$\bar{\bar{\mu}}(\mathbf{r})=\bar{\bar{\mu}}(\mathbf{r})^{\rm T}$], the second and third  terms of \eqnref{eq6} locally cancel out [$\mathbf{H}_2 \bar{\bar{\mu}}(\mathbf{r}) \mathbf{H}_1-\mathbf{H}_1 \bar{\bar{\mu}}(\mathbf{r}) \mathbf{H}_2$=0] and reciprocity holds regardless of the specific arrangement of materials. 
This clear demonstration has counter-intuitive consequences when one considers magnetic metamaterials with cleverly designed anisotropies. Consider for example the example shown in Fig.~\ref{fig1}, consisting of two pieces of anisotropic but symmetric magnetic material. Each piece has an anisotropy axis indicated by the arrow; along this direction the relative permeability is $\mu=6$ and along the perpendicular one $\mu=1/6$. When the source of field (a point dipole) is placed on the left, the magnetic field transferred to the right of the material is shown in (a) [colors represent the $B_z$ component of field]. When this same source is placed on the right, the field transferred to the left side of the material is very different (b). Despite this clear asymmetric field transfer, materials are locally symmetric and reciprocity is clearly fulfilled showing that $B_{z,1}(r_2)=B_{z,2}(r_1)$, where $B_{z,1}$ is the field distribution when the dipole is in position $r_1$ and $B_{z,2}$ when is in position $r_2$.

A second strategy to break magnetic reciprocity involves the use of non-linear magnetic materials, whose permeability depends on the local magnetic field, $\bar{\bar{\mu}}(\mathbf{B})$. Actually, most magnetic materials shown non-linear behaviours for sufficiently large magnetic fields~\cite{Coey}, and therefore, this is a feasible approach to break magnetic reciprocity. However, the use of non-linearities restricts the application to a small range of field amplitudes. Non-linear magnetic materials entail other drawbacks, like remanent magnetizations, which persist even after the external source of field is removed. In addition, the non-reciprocal response cannot be actively controlled since it relies on the intrinsic properties of the material used. 

\subsection{Moving conductor}

 When a \textit{conductive material} (with electrical conductivity $\sigma$) is at rest, the current density appearing in the material fulfills $\mathbf{J}=\sigma \mathbf{E}$. When the conductor moves in presence of an external magnetic field ($K$ is the lab frame and $K'$ is the reference frame in the conductor), the current density appearing in the conductor has a different form~\cite{Landau}
\begin{equation}
\mathbf{J}=\sigma(\mathbf{E}+\mathbf{v}\times\mathbf{B}), \label{mov.conduct}
\end{equation}
where $\mathbf{v}$ is the velocity of the medium (i.e. the velocity of $K'$ respect to $K$) which is assumed to be small compared to the speed of light (non-relativistic~\cite{Jackson}). In this expression the second term shows how the movement of the conductor in presence of the magnetic field creates an effective electric field in the $K'$ reference frame. From the lab frame $K$, the appearance of the this current can be explained via the Lorentz force equation $\mathbf{F}=q(\mathbf{E}+\mathbf{v}\times \mathbf{B})$; charges contained in the conductor experience a force as they move in presence of the external magnetic field which generates a current density in the conductor. 
One can now repeat the development of the magnetostatic reciprocity condition adding the term of a moving conductor [\eqnref{mov.conduct}] into \eqnref{eq1} reading {$\nabla \times \mathbf{H}_j=\mathbf{J}^{\rm f}_j+\sigma \mathbf{v}\times\mathbf{B}_j$ (where we have omitted the electric field, since we only consider the existence of static magnetic fields). Following the same procedure, one finds 
\begin{equation}
\begin{split}
\nabla \cdot ( \mathbf{H}_1 \times \mathbf{A}_2 -\mathbf{H}_2 \times \mathbf{A}_1 )= & \, \mathbf{H}_2 \bar{\bar{\mu}} \mathbf{H}_1-\mathbf{H}_1 \bar{\bar{\mu}} \mathbf{H}_2 \\
&+\sigma \mathbf{v}\cdot[(\nabla \times \mathbf{A}_1)\times \mathbf{A}_2-(\nabla \times \mathbf{A}_2)\times \mathbf{A}_1]\\
&+\mathbf{A}_2\cdot\mathbf{J}^{\rm f}_1-\mathbf{A}_1\cdot\mathbf{J}^{\rm f}_2,
\end{split}
\end{equation}
which substitutes \eqnref{eq6}. 
The third term on the right-hand-side of this equation comes from the movement of the conductor and, in general, is different from zero because $(\nabla \times \mathbf{A}_1)\times \mathbf{A}_2\neq (\nabla \times \mathbf{A}_2)\times \mathbf{A}_1$. This shows how a linear and isotropic electrical conductor moving with constant velocity is able to break magnetic reciprocity. }


\newpage
\section{Magnetic field of a magnetic dipole moving above a conducting half-space}

Here we are going to derive and characterize the magnetic field created by a magnetic dipole moving in front of a conducting half-space. In the lab frame $(\rr,t)$ the magnetic dipole is situated at $z=z_0>0$ and moves with a constant velocity $v$ along the $x$-axis. The conducting half-space extends over $z<0$. The conductive material is modeled as a linear material with a relative frequency-independent permeability $\mu$ and a relative permittivity $\epsilon(\omega)=1+\im \sigma/(\epsilon_0\omega)$, which follows from the Drude model in the quasi-static limit. In \ref{subsec:1} we derive an analytical expression for the magnetic field of a static dipole. In \ref{subsec:2} we derive and discuss the force acting on the dipole. In \ref{subsec:3} we derive the magnetic field for an oscillating dipole.
\subsection{Magnetic field for a static dipole}\label{subsec:1}
We assume that at $t=\tilde t=0$ the lab frame $(\rr,t)$ and the rest frame of the magnetic dipole $(\tilde\rr, \tilde t)$ coincide. Therefore, in its rest frame, the magnetic dipole is situated at $\tilde\rr=z_0\uv_z$. Given that the dipole moment of the magnetic dipole is $\mathbf{m}=m_x \uv_x+m_y\uv_y+m_z\uv_z$, the polarization and magnetization of the dipole reads
\begin{align}
\tilde{\mathbf{P}}(\tilde\rr,\tilde t)&=\mathbf{0},\\
\tilde{\mathbf{M}}(\tilde\rr,\tilde t)&=\mathbf{m}\delta(\tilde x)\delta(\tilde y)\delta(\tilde z-z_0).
\end{align}
The polarization and magnetization in the lab frame can be obtained via Lorentz transformation \cite{Kholmetskii2016}
\begin{align}
\mathbf{P}(\rr,t)&=\frac{\gamma\beta}{c}\delta(\gamma x - \gamma vt)\delta(y)\delta(z-z_0)(-m_z\uv_y+ m_y\uv_z),\\
\mathbf{M}(\rr,t)&=\delta(\gamma x - \gamma vt)\delta(y)\delta(z-z_0)(m_x\uv_x+\gamma m_y \uv_y+\gamma m_z \uv_z),
\end{align}
where, $\gamma\equiv 1/\sqrt{1-\beta^2}$, $\beta=v/c$ and $c$ denotes the speed of light in vacuum. In the spectral domain, using the convention $f(\kk,\omega)=\int_{\mathds{R}^4}\text{d}\rr\,\text{d}t f(\rr,t) \exp[\im (\omega t-\kk\cdot\rr)]$, the polarization and magnetization read
\begin{align}
\mathbf{P}(\rr,\omega)&=\text{sign}(v)\frac{\exp(\im\omega x/v)}{c^2}\delta(y)\delta(z-z_0)(-m_z\uv_y+ m_y\uv_z),\\
\mathbf{M}(\rr,\omega)&=\text{sign}(v)\frac{\exp(\im\omega x/v)}{\gamma v}\delta(y)\delta(z-z_0)(m_x\uv_x+\gamma m_y \uv_y+\gamma m_z \uv_z).
\end{align}
 The current density is given by $\mathbf{J}\coo=-\im\omega\mathbf{P}(\rr,\omega)+\nabla\times\mathbf{M}(\rr,\omega)=\exp(\im\omega x/v)\mathbf{J}(y,z,\omega)$, where
\begin{align}
J_x(y,z,\omega)&=\frac{m_z\partial_y\delta(y)\delta(z-z_0)}{|v|}-\frac{m_y\delta(y)\partial_z\delta(z-z_0)}{|v|},\\
J_y(y,z,\omega)&=-\frac{\im m_z \text{sign}(v)\omega\delta(y)\delta(z-z_0)}{\gamma^2v^2}+\frac{m_x\delta(y)\partial_z\delta(z-z_0)}{\gamma|v|},\\
J_z(y,z,\omega)&=\frac{\im m_y \text{sign}(v) \omega \delta(y)\delta(z-z_0)}{\gamma^2v^2}-\frac{m_x\partial_y\delta(y)\delta(z-z_0)}{\gamma|v|}.
\end{align}
The magnetic field corresponding to this current density distribution is determined by the dyadic Green function
 \begin{equation}
 \mathbf{B}(\rr,\omega)=\mu_0\nabla\times\int_{\mathds{R}^3}\text{d}\rr'\exp(\im \omega x'/v)\mathds{G}(\rr,\rr',\omega)\mathbf{J}(y',z',\omega).
 \end{equation}
For a translational invariant dyadic Green function along $x$ and $y$, i.e. $\mathds{G}(\rr,\rr',\omega)=\mathds{G}(x-x',y-y',z,z',\omega)$ we have that
\begin{equation}
\mathbf{B}(\rr,\omega)=\mu_0\nabla\times\left\lbrace\exp(\im \omega x/v)\int_{\mathds{R}^2}\text{d}y'\text{d}z'\mathds{G}(k_x=\omega/v,y-y',z,z',\omega)\mathbf{J}(y',z',\omega)\right\rbrace.
\end{equation}
And consequently the component $B_i(\rr,\omega)$ of the magnetic field associated to a dipole moment $m_i\uv_i$ is given by
\begin{align}
\label{eq:1}
B_x(\rr,\omega)&=\frac{\mu_0 m_x}{2\pi\gamma |v|}\exp(\im \omega x/v)\int_{\mathds{R}}\text{d}k_y\exp(\im k_y y)\left[\partial_z\partial_{z'}\mathds{G}_{yy}+\im k_y \partial_z \mathds{G}_{yz}-\im k_y \partial_{z'}\mathds{G}_{zy}+k_y^2\mathds{G}_{zz}\right],\\
\label{eq:2}
B_y(\rr,\omega)&=\frac{\mu_0 m_y}{2\pi \gamma^2|v|^3}\exp(\im \omega x/v)\int_{\mathds{R}}\text{d}k_y\exp(\im k_y y)\left[\gamma^2v^2\partial_z\partial_{z'}\mathds{G}_{xx}+\im\omega v \partial_z\mathds{G}_{xz}-\im\gamma^2\omega v \partial_{z'}\mathds{G}_{zx}+\omega^2\mathds{G}_{zz}\right],\\
\label{eq:3}
B_z(
\rr,\omega)&=\frac{\mu_0m_z}{2\pi \gamma^2|v|^3}\exp(\im \omega x/v)\int_{\mathds{R}}\text{d}k_{y}\exp(\im k_y y)\left[\gamma^2v^2k_y^2\mathds{G}_{xx}-vk_y\omega\mathds{G}_{xy}-vk_y\omega\gamma^2 \mathds{G}_{yx}+\omega^2\mathds{G}_{yy}\right],
 \end{align}
where $\mathds{G}_{ij}\equiv \mathds{G}_{ij}(k_x=\omega/v,k_y,z,z_0,\omega)$. We are now going to evaluate the field $B_i(\rr^\pm)=B^0_i(\rr^\pm)+B^s_i(\rr^\pm)$ at $\rr^\pm=(vt\pm \delta>0,0,z_0)$, where $B^0_i$ and $B^s_i$ denotes the free-space and scattering part respectively. The free-space part can easily be obtained via Lorentz transformation, namely
\begin{align}
B_x^0(\rr^\pm)&=\frac{\mu_0m_x}{2\pi\gamma^3z_0^3|\tilde\delta|^3},\\
B_y^0(\rr^\pm)&=-\frac{\mu_0m_y}{4\pi \gamma^2z_0^3|\tilde\delta|^3},\\
B_z^0(\rr^\pm)&=-\frac{\mu_0m_z}{4\pi \gamma^2z_0^3|\tilde\delta|^3},
\end{align}
where $\tilde\delta\equiv\delta/z_0$. The scattering part can be obtained using the scattering dyadic Green function for this setup \cite{Buhmann2012}, which leads to
\begin{align}
B_x^s(\rr^\pm)&=\frac{\mu_0 \gamma^2m_x}{8 \pi^2 z_0^3}\int_0^\infty\int_{-\pi}^{\pi}\text{d}\xi\text{d}\phi\frac{\exp(-2\xi)\xi^2\sin^2(\phi)}{1+\beta^2\gamma^2\sin^2(\phi)}\exp[\pm\im\,\text{sign}(v)\gamma\sin(\phi)\xi\tilde\delta]\left[r_s(\xi,\phi)+\beta^2\cos^2(\phi)r_p(\xi,\phi)\right],\\
B_y^s(\rr^\pm)&=\frac{\mu_0\gamma m_y}{8\pi^2z_0^3}\int_0^\infty\int_{-\pi}^{\pi}\text{d}\xi\text{d}\phi\frac{\exp(-2\xi)\xi^2\cos^2(\phi)}{1+\beta^2\gamma^2\sin^2(\phi)}\exp[\pm\im\,\text{sign}(v)\gamma\sin(\phi)\xi\tilde\delta]\left[r_s(\xi,\phi)-\gamma^2\beta^2\sin^2(\phi)r_p(\xi,\phi)\right],\\
B_z^s(\rr^\pm)&=\frac{\mu_0\gamma m_z}{8\pi^2z_0^3}\int_0^\infty\int_{-\pi}^{\pi}\text{d}\xi\text{d}\phi\exp(-2\xi)\xi^2\exp[\pm\im\,\text{sign}(v)\gamma\sin(\phi)\xi\tilde\delta] r_s(\xi,\phi).
\end{align}
The reflection coefficients are determined by the relative permeability and permittivity and equal
\begin{align}
r_s(\xi,\phi)&=\frac{\mu\sqrt{\xi^2}-\sqrt{\xi^2-[\mu\epsilon(\gamma\omega_c\xi\sin(\phi))-1][\beta\gamma\xi\sin(\phi)]^2}}{\mu\sqrt{\xi^2}+\sqrt{\xi^2-[\mu\epsilon(\gamma\omega_c\xi\sin(\phi))-1][\beta\gamma\xi\sin(\phi)]^2}},\\
r_p(\xi,\phi)&=\frac{\epsilon(\gamma\omega_c\xi\sin(\phi))\sqrt{\xi^2}-\sqrt{\xi^2-[\mu\epsilon(\gamma\omega_c\xi\sin(\phi))-1][\beta\gamma\xi\sin(\phi)]^2}}{\epsilon(\gamma\omega_c\xi\sin(\phi))\sqrt{\xi^2}+\sqrt{\xi^2-[\mu\epsilon(\gamma\omega_c\xi\sin(\phi))-1][\beta\gamma\xi\sin(\phi)]^2}},
\end{align}
where $\omega_c\equiv |v|/z_0$ denotes the characteristic frequency of the system. Note that for a purely real permittivity we have that $r_s(\xi,-\phi)=r_s(\xi,\phi)$ and $r_p(\xi,-\phi)=r_p(\xi,\phi)$, which implies $B_i(\rr^+)=B_i(\rr^-)$. For non-relativistic velocities, i.e. $|\beta|\ll1$ these expressions simplify to
\begin{align}
B_x^s(\rr^\pm)&\approx\frac{\mu_0 m_x}{8 \pi^2 z_0^3}\int_0^\infty\int_{-\pi}^{\pi}\text{d}\xi\text{d}\phi\exp(-2\xi)\xi^2\sin^2(\phi)\exp[\pm\im\,\text{sign}(v)\sin(\phi)\xi\tilde\delta]r_s(\xi,\phi),\\
B_y^s(\rr^\pm)&\approx\frac{\mu_0 m_y}{8\pi^2z_0^3}\int_0^\infty\int_{-\pi}^{\pi}\text{d}\xi\text{d}\phi\exp(-2\xi)\xi^2\cos^2(\phi)\exp[\pm\im\,\text{sign}(v)\sin(\phi)\xi\tilde\delta]r_s(\xi,\phi),\\
B_z^s(\rr^\pm)&\approx\frac{\mu_0 m_z}{8\pi^2z_0^3}\int_0^\infty\int_{-\pi}^{\pi}\text{d}\xi\text{d}\phi\exp(-2\xi)\xi^2\exp[\pm\im\,\text{sign}(v)\sin(\phi)\xi\tilde\delta] r_s(\xi,\phi).
\end{align}
For $\mu=1$ and $\epsilon=1+\im \sigma/(\epsilon_0\omega)$ the reflection coefficient is given by
\begin{align}
r_s(\xi,\phi)&\approx\frac{\sqrt{\xi^2}-\sqrt{\xi^2-\im R_m\xi\sin(\phi)}}{\sqrt{\xi^2}+\sqrt{\xi^2-\im R_m\xi\sin(\phi)}},
\end{align}
where $R_m$ denotes the magnetic Reynolds number and is given by $R_m\equiv\mu_0\sigma |v| z_0$. 
\subsection{Dissipated power}\label{subsec:2}
The force acting on the magnetic dipole is given by $\mathbf{F}=\int_{\mathds{R}^3}\text{d}\rr\,\mathbf{J}(\rr,t)\times\mathbf{B}^s(\rr,t)$. One can easily show that for a point magnetic dipole the force is given by $\mathbf{F}=\sum_i [\mathbf{m}\partial_{x_i}\mathbf{B}^s(vt,0,z_0)]\uv_i$. 
In order to maintain the dipole at a constant velocity one has to apply an equal and opposite force. For $|\beta| \ll 1$ this equals a power $P_i=-vF^i_x$, where 
\begin{align}
F^x_x&=m_x\partial_x B^s_x(v t,0,z_0)=m_x \lim\limits_{\delta\rightarrow 0}\frac{B_x(\rr^+)-B_x(\rr^-)}{2\delta},\\
F^y_x&=m_y\partial_x B^s_y(v t,0,z_0)= m_y \lim\limits_{\delta\rightarrow 0}\frac{B_y(\rr^+)-B_y(\rr^-)}{2\delta},\\
F^z_x&=m_z\partial_x B^s_z(v t,0,z_0)= m_z \lim\limits_{\delta\rightarrow 0}\frac{B_z(\rr^+)-B_z(\rr^-)}{2\delta}.
\end{align}
Let us now define a dimensionless force $\tilde F_x^i= 8\pi^2 z_0^4F_x^i/(\mu_0 m_i^2)$, which is a function of $R_m$ only. One can easily find asymptotic approximations for these expressions, namely
\begin{align}
\tilde{F}_x^x&\approx\text{sign}(v)\begin{cases}
-3\pi R_m/64&\,\,\text{for}\,\,R_m\ll 1,\\
-2.1R_m^{-1/2}&\,\,\text{for}\,\,R_m\gg 1,
\end{cases}\\
\tilde{F}_x^y&\approx\text{sign}(v)\begin{cases}
-\pi R_m/64&\,\,\text{for}\,\,R_m\ll 1,\\
-1.4R_m^{-1/2}&\,\,\text{for}\,\,R_m\gg 1,
\end{cases}\\
\tilde{F}_x^z&\approx\text{sign}(v)\begin{cases}
-4\pi R_m/64&\,\,\text{for}\,\,R_m\ll 1,\\
-3.5R_m^{-1/2}&\,\,\text{for}\,\,R_m\gg 1.
\end{cases}
\end{align}
One can therefore conclude that there exists a $R_m$ for which the power is maximal. Note that one can also express the scattered field for $R_m\ll 1$, namely
\begin{align}
B_x^s(\rr^\pm)&\approx\pm\text{sign}(v)\frac{\mu_0 m_x R_m}{16\pi z_0^3}\frac{1}{\tilde{\delta}^3}\left[4\left(\frac{1}{(4+\tilde{\delta}^2)^{1/2}}-1\right)+4\frac{4+2\tilde{\delta}^2}{(4+\tilde{\delta}^2)^{3/2}}\right],\\
B_y^s(\rr^\pm)&\approx\mp\text{sign}(v)\frac{\mu_0 m_y R_m}{16\pi z_0^3}\frac{1}{\tilde{\delta}^3}\left[4\left(\frac{1}{(4+\tilde{\delta}^2)^{1/2}}-1\right)+(4+\tilde{\delta}^2)^{1/2}\right],\\
B_z^s(\rr^\pm)&\approx \mp\text{sign}(v)\frac{\mu_0 m_z R_m}{16 \pi z_0^3}\frac{\tilde\delta}{(4+\tilde\delta^2)^{3/2}},
\end{align}
where the z-component is equal to the expression obtained in \cite{Thess2007}. 
\subsection{Magnetic field for an oscillating dipole}\label{subsec:3}
Let us now derive the magnetic field for a dipole with a dipole moment $\mathbf{m}=m\uv_z$ oscillating at angular frequency $\omega_0$ with $|\beta|\ll1$ and $\mu=1$. In this case the magnetization reads 
\begin{align}
 \mathbf{M}(\rr,t)&=m_z\cos(\omega_0 t)\delta(x-vt)\delta(y)\delta(z-z_0)\uv_z,\\
 \mathbf{M}(\rr,\omega)&=\frac{m_z}{2|v|}\sum_s\exp(\im\omega_s x/v)\delta(y)\delta(z-z_0)\uv_z,
 \end{align}
 where $\omega_\pm=\omega\pm\omega_0$. The current density is given by $\mathbf{J}\coo=\nabla\times\mathbf{M}(\rr,\omega)=\sum_{l=\pm}\exp(\im\omega_l x/v)\mathbf{J}(y,z,\omega)$, where
 \begin{align}
 j_x(y,z,\omega)&=\frac{m_z\partial_y\delta(y)\delta(z-z_0)}{2|v|},\\
 j_y(y,z,\omega)&=-\frac{\im m_z \text{sign}(v)\omega_s\delta(y)\delta(z-z_0)}{2v^2},\\
 j_z(y,z,\omega)&=0.
 \end{align}
Following the same approach as in the first section, the component $B_z(\rr,\omega)$ of the magnetic field reads
\begin{align}
B_z(\rr,\omega)&=\frac{\mu_0m_z}{4\pi |v|^3}\sum_{l=\pm}\exp(\im \omega_l x/v)\int_{\mathds{R}}\text{d}k_{y}\exp(\im k_y y)\left[v^2k_y^2\mathds{G}_{xx}-vk_y\omega_l\mathds{G}_{xy}-vk_y\omega_l \mathds{G}_{yx}+\omega_l^2\mathds{G}_{yy}\right],
\end{align}
where $\mathds{G}_{ij}\equiv \mathds{G}_{ij}(k_x=\omega_l/v,k_y,z,z_0,\omega)$. The free space magnetic field component is given by $B_z^0(\rr^\pm,t)=B^0_{\text{in}}\cos(\omega_0t)+B^0_{\text{out}}\sin(\omega_0 t)$, where
\begin{align}
B^0_\text{in}&=\frac{\mu_0m_z}{4\pi|\delta|^3}\left[[(\beta|\tilde\delta|\omega_0/\omega_c)^2-1]\cos(\beta|\tilde\delta|\omega_0/\omega_c)-(\beta|\tilde\delta|\omega_0/\omega_c)\sin(\beta|\tilde\delta|\omega_0/\omega_c)\right],\\
B^0_\text{out}&=\frac{\mu_0m_z}{4\pi|\delta|^3}\left[[\beta|\tilde\delta|\omega_0/\omega_c)^2-1]\sin(\beta|\tilde\delta|\omega_0/\omega_c)+(\beta|\tilde\delta|\omega_0/\omega_c)\cos(\beta|\tilde\delta|\omega_0/\omega_c)\right].
\end{align}
Let us now evaluate the scattering component of the magnetic field. According to the scattering dyadic Green function for this setup \cite{Buhmann2012} we get
\begin{align}
B_z^s(\rr^\pm,t)&=\frac{\mu_0m_z}{16\pi^2 z_0^3}\int_0^\infty\int_{-\pi}^{\pi}\text{d}\xi\text{d}\phi\exp(-2\xi)\xi^2\exp(\im\omega_0 t)\exp[\pm\im\,\text{sign}(v)\xi\sin(\phi)\tilde\delta]r_{s+}(\xi,\phi)\\
&+\frac{\mu_0m_z}{16\pi^2 z_0^3}\int_0^\infty\int_{-\pi}^{\pi}\text{d}\xi\text{d}\phi\exp(-2\xi)\xi^2\exp(-\im\omega_0 t)\exp[\pm\im\,\text{sign}(v)\xi\sin(\phi)\tilde\delta]r_{s-}(\xi,\phi),
\end{align}
where the reflection coefficients are given by
\begin{equation}
r_{s\pm}(\xi,\phi)=\frac{\sqrt{\xi^2}-\sqrt{\xi^2-\im R_m[\xi\sin(\phi)\mp\omega_0/\omega_c]}}{\sqrt{\xi^2}+\sqrt{\xi^2-\im R_m[\xi\sin(\phi)\mp\omega_0/\omega_c]}}.
\end{equation}
Therefore one can express the scattering component analogously in terms of an in-phase and out-of-phase component $B_z^s(\rr^\pm)=B^s_\text{in}\cos(\omega_0t)+B^s_\text{out}\sin(\omega_0 t)$, where
\begin{align}
B^s_\text{in}&=\frac{\mu_0m_z}{16\pi^2z_0^3}\int_0^\infty\int_{-\pi}^{\pi}\text{d}\xi\text{d}\phi\exp(-2\xi)\xi^2\exp[\pm\im\,\text{sign}(v)\xi\sin(\phi)\tilde\delta][r_{s+}(\xi,\phi)+r_{s-}(\xi,\phi)],\\
B^s_\text{out}&=\frac{\im \mu_0m_z}{16\pi^2z_0^3}\int_0^\infty\int_{-\pi}^{\pi}\text{d}\xi\text{d}\phi\exp(-2\xi)\xi^2\exp[\pm\im\,\text{sign}(v)\xi\sin(\phi)\tilde\delta][ r_{s+}(\xi,\phi)- r_{s-}(\xi,\phi)].
\end{align}
Let us now consider the regime where $\omega_0/\omega_c=\omega_0z_0/|v|\ll 1$. First of all note that in this regime we have that
\begin{equation}
B_z^0(\rr^\pm,t)\approx B_z^0(\rr^\pm)\cos(\omega_0 t).
\end{equation}
Moreover we have that $r_{s+}(\xi,\phi)\approx r_{s-}(\xi,\phi)$ and therefore
\begin{equation}
B_z^s(\rr^\pm,t)\approx B_z^s(\rr^\pm)\cos(\omega_0 t).
\end{equation}
Therefore we can conclude that for $\omega_0/\omega_c\ll1$ the the magnetic field is given by the magnetic field of the static case modulated by $\cos(\omega_0 t)$.

{
\subsection{Dynamics of electrons exposed to electromagnetic fields in a conductor}
In this subsection of the supplementary material we partially follow the derivation of the Drude-model in \cite{Ashcroft}. First, we assume that the velocity of an electron in a conductor can only be changed by collisions with impenetrable ion cores and external electromagnetic fields. The collision results in an instantaneous change in velocity. After each collision an electron is taken to emerge with a randomly directed velocity of temperature dependent absolute value. The probability that an electron experiences a collision per unit time is assumed to be constant and given by $1/\tau$. For copper at room temperature we have $\tau\approx 2.49\times 10^{-14}\text{s}$. The typical mean free path an electron travels between collisions at room temperature is on the order of a few hundreds of $\AA$ngstr\"om. Let us now define the average momentum per electron at any $\rr$ and $t$ by $\mathbf{p}(\rr,t)$. The current density is related to the average momentum via $\mathbf{j}(\rr,t)=-en\mathbf{p}(\rr,t)/m_e$, where $-e\approx -1.6\times 10^{-19}\text{C}$ denotes the electron charge, $n\,(\approx 8.47\times 10^{28}/\text{m}^3\text{  for copper at room temperature})$ denotes the free-electron density and $m_e\approx 9.1\times 10^{-31}\text{kg}$ denotes the electron mass. Assuming that the external electromagnetic fields do not vary appreciably over the length scale of several mean free paths we have that after some time $\delta t$ the average momentum per electron is given by
\begin{align}
\mathbf{p}(\rr,t+\delta t)&=\left(1-\frac{\delta t}{\tau}\right)\left[\mathbf{p}(\rr,t)-e\mathbf{E}(\rr,t)\delta t-\frac{e}{m_e}\mathbf{p}(\rr,t)\times\mathbf{B}(\rr,t) \delta t+\mathcal{O}(\delta t)^2\right]+\mathcal{O}(\delta t)^2\\
&=\mathbf{p}(\rr,t)-\frac{\delta t}{\tau}\mathbf{p}(\rr,t)-e\mathbf{E}(\rr,t)\delta t-\frac{e}{m_e}\mathbf{p}(\rr,t)\times\mathbf{B}(\rr,t)\delta t+\mathcal{O}(\delta t)^2
\end{align}
We may therefore write in the limit $\delta t\rightarrow 0$
\begin{align}
\left(\tau\frac{\text{d}}{\text{d}t}+1\right)\mathbf{j}(\rr,t)&=\frac{e^2 n  \tau}{m_e}\left[\mathbf{E}(\rr,t)+\mathbf{v}(\rr,t)\times\mathbf{B}(\rr,t)\right]\\
&=\sigma\mathbf{E}(\rr,t)-\frac{\sigma}{en}\mathbf{j}(\rr,t)\times\mathbf{B}(\rr,t),
\end{align}
where $\sigma\,(\approx 5.96\times10^7\text{S}/\text{m}\,\text{ for copper at room temperature})$ denotes the static conductivity of copper at room temperature. In the quasi-static limit where $|\tau\,\text{d}\mathbf{j}(\rr,t)/\text{d}t|\ll \mathbf{j}(\rr,t)$ we have that
\begin{equation}
\mathbf{j}(\rr,t)=\sigma \mathbf{E}(\rr,t)-\frac{\sigma}{en}\mathbf{j}(\rr,t)\times\mathbf{B}(\rr,t).
\end{equation}
Solving this equation for the current density leads to
\begin{equation}
\mathbf{j}(\rr,t)=\frac{\sigma }{1+(\sigma/en)^2\mathbf{B}^2(\rr,t)}\left[\mathbf{E}(\rr,t)-(\sigma/en)\mathbf{E}(\rr,t)\times \mathbf{B}(\rr,t)+ (\sigma/en)^2 (\mathbf{E}(\rr,t)\cdot\mathbf{B}(\rr,t))\mathbf{B}(\rr,t)\right].
\end{equation}
So for small enough magnetic fields, i.e. $||\sigma/(en)\mathbf{B}(\rr,t)||\approx 4.3\times10^{-3}||\mathbf{B}(\rr,t)||\text{T}^{-1}\ll 1$ for copper, this expression is approximately equal to
\begin{equation}
\mathbf{j}(\rr,t)\approx\sigma[\mathbf{E}(\rr,t)-(\sigma/en)\mathbf{E}(\rr,t)\times\mathbf{B}(\rr,t)].\label{eqa}
\end{equation}
In this equation, the first term is the linear Ohmic relation between electric field and induced current density. The second term is similar to the expression of the current density induced in a moving conductor   discussed in the main text. Actually, if one considers the dipole's rest frame, the conducting half space moves at a velocity $-v$ along the $x$-axis. The electric field in the lab frame can therefore be expressed in the rest frame via a Lorentz transformation $\mathbf{E}(\rr,t)=-\gamma v\uv_x \times \mathbf{B}'(\rr',t')$, since there is no electric field in the rest frame. For small velocities $|\beta|\ll 1$, Maxwell equations in the rest frame then read
\begin{align}
\nabla\cdot \mathbf{B}(\rr)&=0,\\
\nabla\times\mathbf{B}(\rr)&=\mu_0[\nabla\times \mathbf{M}(\rr)-\Theta(-z)\sigma v\uv_x\times \mathbf{B}(\rr)].\label{eqb}
\end{align}
By comparing equations (\ref{eqa}) and (\ref{eqb}) one can see how the application of an electric field gives rise to current density term [second term in the right-hand-side of Eq. (\ref{eqa})] that is analogous to the current density appearing in a moving conductor in the rest frame [second term in the right-hand-side of Eq. (\ref{eqb})]. 
By equating these two current densities, one finds that the electric field $E_0$ required to generate the same current density is given by $E_0 = env/\sigma=v/\mu_e$, where $\mu_e$ denotes the electron mobility.

For copper we have $\mu_e\approx 0.0044$m$^2$/(Vs), such that for a linear velocity of $v=3$m/s the required electric field would be $E_0\approx 682V/m$. This would result in a current density $J=\sigma E_0\approx 4\times 10^{10}$A/m$^2$, which is more than three orders of magnitude bigger than the standard maximum current density for copper of $5\times10^6$A/m$^2$. 

}

\newpage
\section{Experimental realization}

\begin{figure}[htb]
\begin{center}
\includegraphics[width=1.0\textwidth]{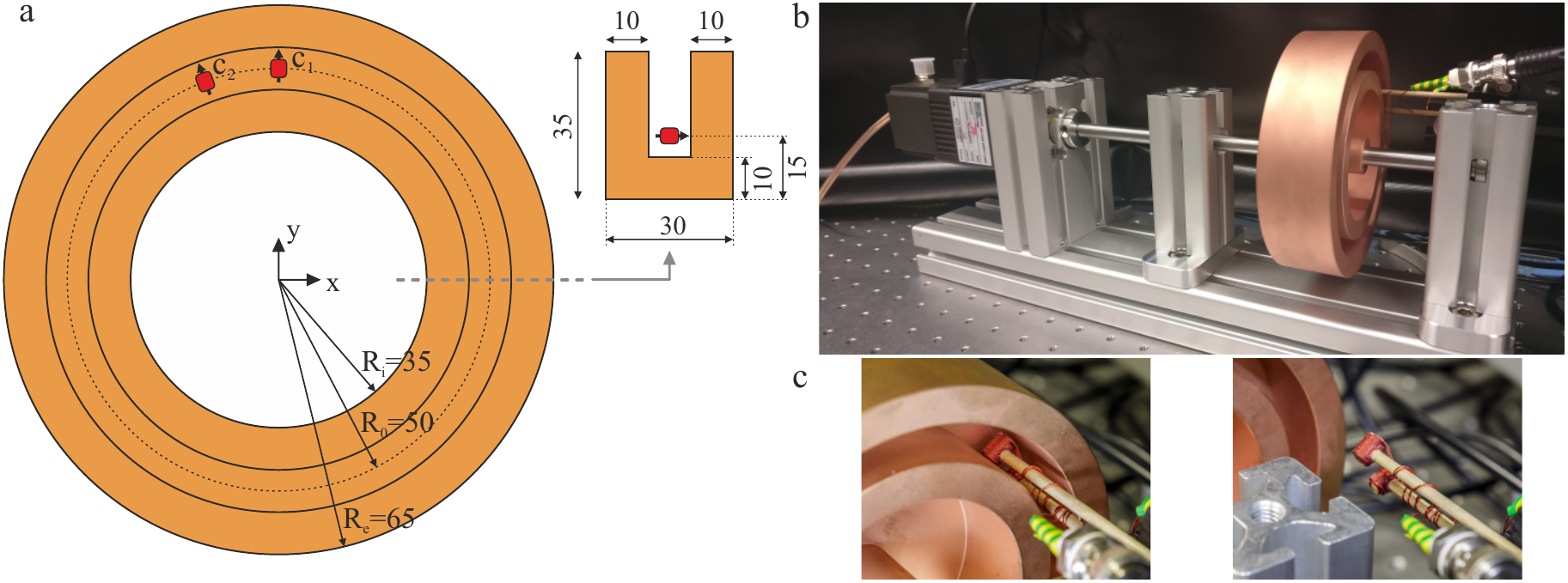}
\caption{(a) Sketch of the conductive part, sizes are given in millimeters. The coils are sketched in red. (b) Picture of the actual setup. {(c) Close-up pictures showing the coils in their original positions for the experiment (left) and after moving the conducting part away from the coils (right). Coils were mounted on wooden sticks to minimize any distortion of the magnetic field.}}
\label{fig2}
\end{center}
\end{figure}

A circularly symmetric piece of electrically conductive material was designed to demonstrate our theoretical results. The conductive piece had an external radius of $R_e=65$mm and a U-shaped cross-section as is detailed in Fig.~\ref{fig2}a. Coils were placed at a radius $R_0=50$mm. The piece was machined from a single block of copper. 

The conductive piece was attached to a shaft, in turn connected to an electric motor, see Fig.~\ref{fig2}b. The motion of the motor was accurately computer-controlled, with a rotation frequency uncertainty of $\approx \pm 0.03$Hz. The whole system was firmly attached to an optical table. The two coils were hold at the appropriate positions by means of two independent structures that were also attached to the optical table. Coils were wound around a non-magnetic cylindrical core and consisted of 4 layers with 20 turns each. The final external radius of the coils was $\approx 3$mm with a length of $\approx6$mm {(see Fig.~\ref{fig2}c)}. 

One of the two coils was kept at the same position throughout the entire experiment. This coil ($c_1$) was connected to a signal generator and fed with a sinusoidal signal [$\propto \cos(\omega_0 t)$]. The other coil ($c_2$) was moved to different distances from the first as discussed in the main text ($r_1$, $r_2$, and $r_3$ for which the distances from center-to-center of coils were 11.4, 13.1, and 15.5mm, respectively) and was connected to a lock-in amplifier (Stanford Research Systems Model SR830 DSP). The two components of the voltage induced in the pick-up coil were recorded, being  $V^{x}$ the part of the voltage in-phase with the generated signal [$\propto \cos(\omega_0 t)$], and $V^{y}$ the out-of-phase part of the voltage [$\propto \sin(\omega_0 t)$].

Voltage measurements shown in the main text and in Fig.~\ref{fig3} were normalized to the voltage measured between the two bare coils in free space (which only has $V^{y}$ component) at the same distance, $|V_0(r_i)|$.

\subsection{Finite-element calculations}

We performed 3D numerical calculations of the moving conductor using COMSOL Multiphysics (Magnetic and Electric fields module). The material was characterized by a constant electrical conductivity $\sigma=5.96\times 10^7\Omega^{-1}{\rm m}^{-1}$ and we included the Lorentz term, $\mathbf{J}_{\rm mc}=\sigma \mathbf{v}\times \mathbf{B}$, with $\mathbf{v}=\rho 2\pi \, \nu \hat{\rm e}_{\varphi}$ (being $\rho$, $\varphi$, $z$ the standard cylindrical coordinates). The source coil was represented by a point dipole and the $B_{\rho}$ component of the field was evaluated at different positions in the dipole plane ($z=15$mm).

First, we solved the stationary problem for a static magnetic dipole. In Fig.~\ref{fig3} we include, in the upper plot, the corresponding numerical calculations for the static case in black lines (which are mostly overlapping with the solid color lines).

We then solved the problem in the frequency domain. In this case, Maxwell equation were solved in the spectral representation, having all magnitudes a time-dependence $\propto e^{\im \omega_0 t}$. In the calculations, fields are treated as complex magnitudes, $\mathbf{B}=\mathbf{B}^{r}+\im \mathbf{B}^{i}$, such that $\mathbf{B}(t)=\mathbf{B}^{r} \cos(\omega_0 t)-\mathbf{B}^{i} \sin(\omega_0 t)$ ($\mathbf{B}^{r}$ is the field component in-phase with the source and $-\mathbf{B}^{i}$ the out-of-phase component). Numerical calculations shown in Fig.~3c of the main text and in Fig.~\ref{fig3} were normalized to field calculated in free space (having only $B_{\rho}^{r}$ component) at the same distance, $|B_0(r_i)|$. Color plots in Fig.~3b of the main text correspond to plots of $B_{\rho}^{r}$ at the plane of the dipole ($z=15$mm). 

The shadow areas in the plot of Fig.~3c of the main text were obtained from different numerical calculations in which the physical parameters of the calculations were slightly modified. In this way we were able to account for the uncertainty in the measurement of the actual distance between the coils and the uncertainty on their relative position respect to the conductor. In particular, we considered distances between coils $\pm0.5$mm the nominal value and uncertainties in the z-position of the two coils of $\pm0.5$mm. Shadow areas were defined by the farthest values obtained for each $\nu$. 


\subsection{Measurements of mutual inductance}

The measurement of the mutual inductance between the coils was done in the following way. The frequency of the signal generator was set to $\omega_0/(2\pi)=9$Hz and the coil $c_2$ was placed at position $r_2$. In the original configuration of the setup (config. I), with the coil $c_1$ connected to the signal generator and the coil $c_2$ connected to the lock-in amplifier, we measured the lock-in voltage in free space, in absence of any material near the coils. 
Next, we placed the coils in their appropriate positions near the moving conductive piece. With the same configuration (config. I) we measured the lock-in voltage at zero velocity of the conductor, $\nu=0$. We next connected the coil $c_2$ to the signal generator and $c_1$ to the lock-in (config. II) and we measured the voltage again. 
Finally, we repeated the measurements (first in config. I and then in config. II) for a velocity of the conductor of $\nu=33.3$Hz.  

\begin{figure}[htb]
\begin{center}
\includegraphics[width=0.5\textwidth]{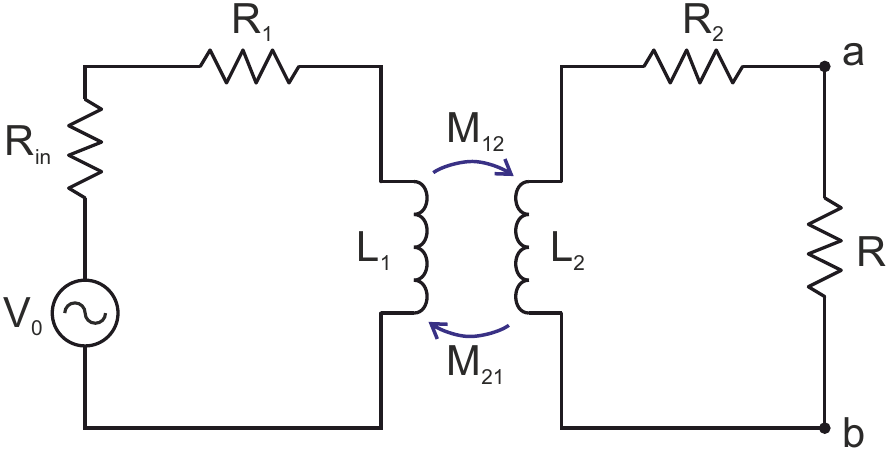}
\caption{Diagram of the circuit used to measure the mutual inductance between coils.}
\label{fig4}
\end{center}
\end{figure}

With these measurements we calculated the mutual inductance between the coils. The electrical scheme of the circuit we used is represented in Fig.~\ref{fig4}. In this circuit $R_{\rm in}$ represents the internal resistance of the signal generator ($V_0=10.6$V, $R_{\rm in}=50\Omega$), $L_1$ and $L_2$ are the self-inductances of the coils and $R_1$ and $R_2$ their corresponding resistances. The internal impedance of the lock-in amplifier is represented by $R$, being the voltage between points a and b ($V_{ab}$) the measured lock-in voltage. The circuit can be easily solved in the frequency domain and, by assuming that $R \rightarrow \infty$ (the input impedance of the lock-in is 10M$\Omega$), the voltage $V_{ab}$ reads $V_{ab}=(-\im \omega_0 M_{12} V_0)/(R_{\rm in}+R_1+\im \omega_0 L_1)$. Considering that $R_{\rm in}\gg |R_1+\im \omega_0 L_1|$, we find 
\begin{equation}
    M_{12}=\frac{R_{\rm in} V_{ab}}{-\im \omega_0 V_0}. \label{eq.M}
\end{equation}

By applying Eq.~\ref{eq.M} to the different measured voltages ($V=V_x+\im V_y$), we calculated the value of mutual inductance for each case, which are summarized in Table~\ref{table1}.
In free space $M$ is purely real. When coils are placed near the conductor at zero velocity, $M$ experiences a slight increase due to the magnetic field expulsion of the conductor (the field becomes slightly concentrated between the conductive parts). In addition, it gets an imaginary part related to the eddy-current losses appearing in the conductor. In any case, for $\nu=0$Hz, the values of $M$ for configurations I and II are the same within the error, in agreement with the reciprocity principle. When the conductor moves at $\nu=33.3$Hz, the mutual inductances for configurations I and II become extremely different, close to zero for config. I and very large for config. II (even larger than the value for $\nu=0$). These measurements demonstrate that the mutual inductance from coil $c_1$ to coil $c_2$ (referred as $M_{12}$ in the main text) is practically cancelled by the moving conductor whilst the mutual inductance from coil $c_2$ to $c_1$ ($M_{21}$ in the main text) is enhanced by it.
As we already pointed out, the imaginary part of the mutual inductance results from the eddy-current losses in the conductor, which can be reduced by decreasing the frequency of the signal, $\omega_0$. In the strict static case inductances would be purely real.  

\begin{table}[h]
\centering
  \caption{Measured mutual inductances (in units of nH).}
  \label{table1}
  \begin{tabular}{ l | c | r }
      & $M$ & $\Delta M$\\
    \hline \hline
    free space, config. I & 21.0 & 0.1 \\ \hline
    $\nu=0$Hz, config. I & $21.5+\im2.7$ & 0.7 \\ \hline
    $\nu=0$Hz, config. II & $21.5+\im3.0$ & 0.7 \\ \hline
    $\nu=33.3$Hz, config. I & $0.0+\im1.9$ & 0.6 \\ \hline
    $\nu=33.3$Hz, config. II & $35.5-\im0.2$ & 0.6 \\ \hline
    \hline
  \end{tabular} 
\end{table}


\subsection{Results for different frequencies}

Here we present the complete set of measurements together with their corresponding numerical calculations. Voltage measurements were obtained for three different signal frequencies; 9, 30, and 65Hz. 

\begin{figure}[t!]
\begin{center}
\includegraphics[width=0.8\textwidth]{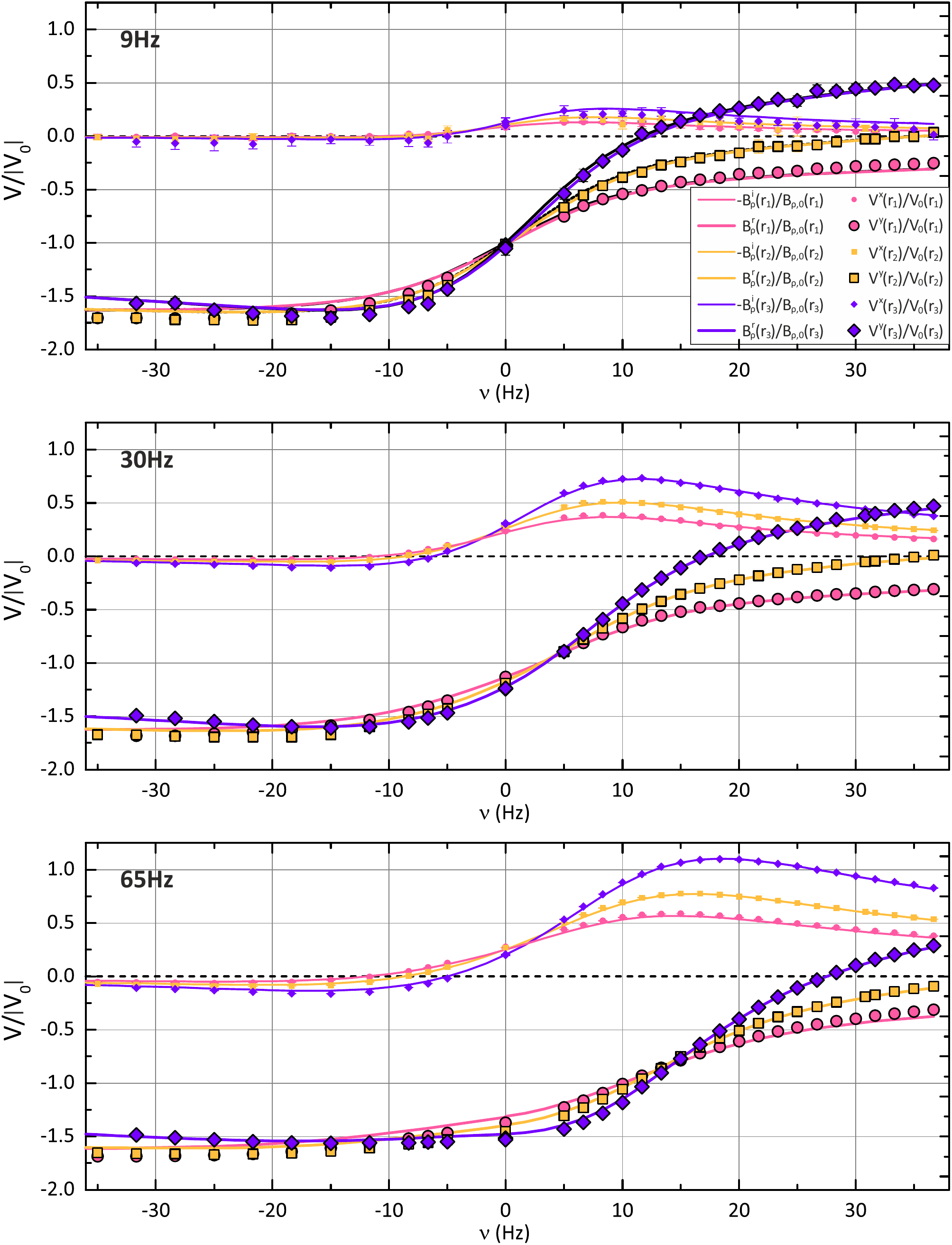}
\caption{Measurements for different frequencies (symbols) and the corresponding numerical calculations (solid lines). Purple, yellow, and pink colors correspond to measurements at positions $r_1$, $r_2$, and $r_3$, respectively. In the 9Hz plot (top) we also included the corresponding numerical calculations for the static case in black lines (which are mostly overlapping with the corresponding solid color lines). {Error bars (1 sigma) are about symbol-size or explictly depicted.}}
\label{fig3}
\end{center}
\end{figure}

The top plot of Fig.~\ref{fig3} corresponds to the measurements shown in Fig.~3 of the main text, with the addition of the measured in-phase voltages ($V^x$) and their corresponding numerical calculations ($-B_{\rho}^i$). These measurements were not included in the main text for the sake of clarity; values are very small and could be further reduced by decreasing the frequency of the signal (they would be exactly zero for static case). In this top plot we also added the numerical calculations for the case of a static dipole (in black lines, solid for $r_3$, dotted for $r_2$, and dashed for $r_1$). As can be seen, static calculations are mostly overlapping with the corresponding calculations for 9Hz, demonstrating that our lock-in measurements for 9Hz reproduce, with good accuracy, the magnetostatic case.

The other two panels for 30 and 65Hz also show a very good agreement with the corresponding numerical calculations. We can observe that, by increasing the signal frequency, measurements move away from the static-low frequency results (in-phase voltages rapidly increase, and out-of-phase voltages change their distribution as well). These results, however, are in agreement with our theoretical analysis. For oscillating dipoles, we showed that one would recover the magnetostatic case if $\omega_0 \ll |v/z_0|$. Considering that $v\approx 2\pi\nu 50\times 10^{-3}$ and $z_0\approx 5\times 10^{-3}$, the "static" approximation should hold for $\nu\gg 3$Hz for a 30Hz signal (and for $\nu\gg 6.5$Hz for a 65Hz signal). As can be seen from the 30Hz plot, the measurements for large (positive and negative) values of $\nu$ agree well with the 9Hz measurements. In contrast, for smaller $\nu$'s, the disagreement between 30Hz and 9Hz measurements is more important. 

